# Rotationally-resolved spectroscopy of Jupiter Trojans (624) Hektor and (911) Agamemnon

D. Perna[(1,2)][1], N. Bott[(2)], T. Hromakina[(2,3)], E. Mazzotta Epifani[(1)], E. Dotto[(1)], A. Doressoundiram[(2)]

[(1)] INAF – Osservatorio Astronomico di Roma, Via Frascati 33, 00078 Monte Porzio Catone (Roma), Italy

[(2)] LESIA – Observatoire de Paris, PSL Research University, CNRS, Sorbonne Universités, UPMC Univ. Paris 06, Univ. Paris Diderot, Sorbonne Paris Cité, 5 place Jules Janssen, 92195 Meudon, France

[(3)] Institute of Astronomy, Kharkiv V.N. Karazin National University, Sumska Str. 35, Kharkiv 61022, Ukraine

**Abstract**

We present the first-ever rotationally resolved spectroscopic investigation of (624) Hektor and (911) Agamemnon, the two largest Jupiter Trojans. The visible and near-infrared spectra that we have obtained at the TNG telescope (La Palma, Spain) do not show any feature or hints of heterogeneity. In particular we found no hints of water-related absorptions. No cometary activity was detected down to ~23.5 R-mag/arcsec$^2$ based on the complementary photometric data. We estimated upper limits on the dust production rates of Hektor and Agamemnon to be ≈30 kg/s and ≈24 kg/s, respectively. We modelled complete visible and near-infrared spectra of our targets using the Shkuratov formalism, to define the upper limit to the presence of water ice and more in general to constrain their surface composition. For both objects, successful models include amorphous carbon, magnesium-rich pyroxene and kerogen, with an upper limit to the amount of water ice of a few percent.

---

[1] E-mail: davide.perna@oa-roma.inaf.it ; davide.perna@obspm.fr

# 1. Introduction

Jupiter Trojans (hereafter: JTs) are primitive asteroids that librate around Jupiter Lagrangian points L4 and L5. Origin and composition of these objects are still matters of debate. At first, they were supposed to have formed in the middle of the solar nebula, near where they currently reside (see Marzari, Tricarico & Scholl 2003, and references therein), but such origin would be difficult to reconcile with the current orbital properties of high-inclination JTs. More recently, Morbidelli et al. (2005) suggested that JTs formed much farther out in the planetesimal disk, and that they have been captured in the region where they currently lie following the gas giants migrations. Such scenario, developed in the framework of the so-called "Nice Model", predicts that JTs represent the most accessible sample of Kuiper belt material (though the physical properties of JTs should be considered more evolved due to such inward migration). The hypothesis that JTs are captured objects from the transneptunian disk has been reinforced by the latest dynamical models (Nesvorný, Vokrouhlický & Morbidelli 2013) and is supported by the similar size distribution of JTs and hot classical transneptunian objects (Fraser et al. 2014). Although the scenario of JTs' formation has not been definitively assessed (see Dotto et al. 2008 and Emery et al. 2015 for reviews of their dynamical and physical properties), it is very plausible to assume that these bodies formed at large heliocentric distances, in volatile- and organic-rich regions. As a consequence, they are probably constituted by silicates and organic compounds, and possibly still contain ices in their interior.

However, visible and near-infrared investigation of JTs (e.g., Dotto et al. 2006; Fornasier et al. 2007) showed featureless spectra, with mean colours very similar to those of short period comets and neutral/blue Centaurs. In particular, no signatures of water ice have been found until now. The current paradigm is that the ice-rich interior of JTs is hidden by a refractory crust, produced over the solar system history by a steady space weathering. In such scenario, occasional impacts could expose the underlying water ice. The most recent models on survival of water ice in JTs (Guilbert-Lepoutre 2014) assess that polar regions can remain cold enough to sustain water ice. Simulations show however that less than 10% of water ice could have survived at most, which would make it very difficult to detect from ground-based observations. On average, water ice should be found ∼10 m below the surface, and possibly below 10 cm in the polar regions.

More recently, Emery, Burr & Cruikshank (2011) identified two spectral groups both in the L4 and L5 swarms. This dichotomy could suggest that we are presently observing the superposition of two distinct populations: the spectrally redder objects (consistent with the asteroidal D-type taxonomic type) would have originated in the trans-Neptunian regions, while the less red group (consistent with the asteroidal P-type classification) would have originated near Jupiter or in the main belt. Until now, no definitive correlations between spectral slope and any other physical or orbital parameter have been evidenced to support the hypothesis that the two spectral groups indeed represent objects with different intrinsic compositions, hence formation regions. However, Brown (2016) reported the presence of absorptions centred around 3.1 μm – possibly related to N-H stretch features – in the less-red population (though the 3-4 μm spectra of Jupiter Trojans form more of a continuum than a dichotomy), and of an unusual deep absorption between 4 and 5 μm in both populations (suggesting a similar composition). A confirmation of these results could support the hypothesis by Wong & Brown (2016) that JTs formed in the outer solar system region straddling the $H_2S$ ice evaporation line. In such scenario, the depletion or retention of

H$_2$S (whose presence on the surface would favour a strong spectral reddening upon irradiation) would be the key factor in creating the current colour bimodality.

The D-types (624) Hektor and (911) Agamemnon (both located in the leading Lagrangian point L4) are among the very largest Jupiter Trojans, which makes them relatively easy targets to investigate. Belonging to the above-mentioned redder spectral group, they have probably originated in the outer regions of the planetesimal disk. Emissivity spectra in the mid-infrared of both Hektor and Agamemnon (Emery, Cruikshank & Van Cleve 2006; Kelley et al. 2017) show features strikingly similar to those of cometary comae, suggesting that they have the same compositions and similar origins in the early solar system. Hektor is a bilobed-shaped body or a contact binary with an equivalent diameter of ~250 km, and a satellite. Its inferred bulk density of 1.0±0.3 g cm$^{-3}$ suggests a high volatile content and/or a high porosity (Marchis et al. 2014). A collisional dynamical family associated with Hektor has been recently attributed by Rozehnal et al. (2016) to a cratering event happened 1-4 Gyr ago. Its rotational period is ~6.9225 hours (Lacerda & Jewitt 2007), while the rotational period of Agamemnon is 6.5819±0.0007 hours (Mottola et al. 2011).

A small bunch of visible spectra exist in the literature for both Hektor (e.g., Cruikshank et al. 2001, and references therein) and Agamemnon (e.g., Lazzaro et al. 2004), while only a few sparse spectral observations have been performed in the near-infrared (NIR). In particular, NIR spectra of Hektor (Dumas, Owen & Barucci 1998; Emery & Brown 2003) seem to be featureless, but differences in the spectral slope suggest some heterogeneity in the composition of the surface. For Agamemnon, Emery (2002) identified a weak feature at 1.74 μm, possibly due to the C-H bond in hydrocarbons, but this result has not been confirmed by subsequent observations (Yang & Jewitt 2007). For both Hektor and Agamemnon, the explanation for the observed variability of the spectra could lie in the fact that different observers were looking (years apart from each other) at different parts of the surfaces, with different compositions. One of the best ways to test this hypothesis is to retrieve rotationally resolved spectroscopy of these bodies, using a single telescope/detector system in order to rule out any systematic effect when comparing the spectra.

In this paper, we report about new visible and NIR spectral observations of Hektor and Agamemnon, obtained at different rotational phases.

## 2. Observations and data reduction

The visible and NIR spectra presented in this work have been obtained at the 3.6-m Telescopio Nazionale Galileo (La Palma, Spain) on two nights, from 18$^{th}$ to 20$^{th}$ December 2013. We used the DOLORES instrument and the LR-B grism to perform visible spectroscopy, while NIR data were obtained with the NICS instrument and the Amici prism. In both cases, the 2 arcsec slit was oriented along the parallactic angle to minimize the risk of flux loss due to the differential refraction. As usual at NIR wavelengths, we used the nodding technique between two positions A and B along the slit: each of the spectra obtained with NICS derives from a complete ABBA sequence. Table 1 reports the observational circumstances.

We used the ESO-MIDAS software package and standard procedures to reduce the obtained data (see Perna et al. 2015 for details). The wavelength calibration of DOLORES data was performed using Ar, Ne+Hg, and Kr lamps' spectral lines, while NIR spectra were calibrated through a look-up table available at the NICS website. The reflectance spectra of Hektor and Agamemnon (presented in Figs. 1 and 2, respectively) were obtained by dividing their visible (four for each target) and NIR (two for each target) spectra by those of well proven solar analog stars (close in observing time and airmass), as reported in Table 1.

Complementary R-band photometric data were acquired on the first observing night (3 images for each target, with an exposure time of 30 s each, yielding a signal-to-noise ratio SNR of ~150-200 at peak, and ~1000 over the PSF area). The images were reduced using standard procedures with the ESO-MIDAS software package: bias subtraction, flat-field correction and aperture photometry to measure the instrumental magnitudes, whose absolute calibration was obtained by observing several Landolt (1992) standard fields.

*Table 1. Observational circumstances.*

| Object | Spec. ID (range) | Date / UT$_{start}$ | T$_{exp}$ (s) | Airmass | Solar Analog (Airmass) | Rot. phase | Spec. slope (%/100 nm) |
|---|---|---|---|---|---|---|---|
| Hektor | 1 (VIS) | 19 Jan 2013 01:02 | 600 | 1.36 | HIP 67460 (1.35) | 0.00 | 12.1±2.0 |
| | 2 (VIS) | 19 Jan 2013 01:41 | 600 | 1.21 | HIP 44027 (1.18) | 0.09 | 11.9±1.9 |
| | 3 (VIS) | 19 Jan 2013 02:22 | 600 | 1.11 | SA98-978 (1.16) | 0.19 | 11.2±1.8 |
| | 4 (VIS) | 19 Jan 2013 04:29 | 600 | 1.04 | HIP 67460 (1.03) | 0.50 | 11.3±1.9 |
| | 5 (NIR) | 20 Jan 2013 02:45 | 4×180 | 1.06 | HIP 44027 (1.30) | 0.72 | 39.4±2.1/10.2±1.9 |
| | 6 (NIR) | 20 Jan 2013 05:31 | 4×120 | 1.14 | 102-1081 (1.16) | 0.11 | 42.5±2.0/11.9±1.8 |
| Agamemnon | 1 (VIS) | 19 Jan 2013 01:24 | 600 | 1.27 | SA102-1081 (1.27) | 0.00 | 10.8±2.0 |
| | 2 (VIS) | 19 Jan 2013 02:00 | 600 | 1.16 | HYADES 64 (1.08) | 0.09 | 10.8±2.1 |
| | 3 (VIS) | 19 Jan 2013 02:42 | 600 | 1.08 | HYADES 64 (1.08) | 0.20 | 10.5±2.0 |
| | 4 (VIS) | 19 Jan 2013 04:50 | 600 | 1.07 | HYADES 64 (1.08) | 0.52 | 10.0±2.1 |
| | 5 (NIR) | 20 Jan 2013 03:12 | 4×180 | 1.05 | HIP 44027 (1.30) | 0.92 | 40.3±2.1/14.4±1.9 |
| | 6 (NIR) | 20 Jan 2013 05:51 | 4×60 | 1.19 | 102-1081 (1.16) | 0.32 | 42.9±1.9/15.9±1.9 |

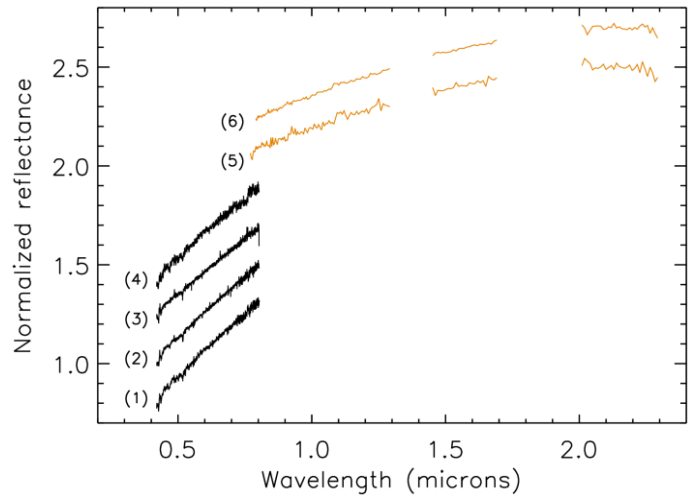

Fig. 1. Reflectance spectra of Hektor, shifted for clarity. Regions affected by strong atmospheric absorption (at ≈1.4 µm and ≈1.9 µm) have been cut out.

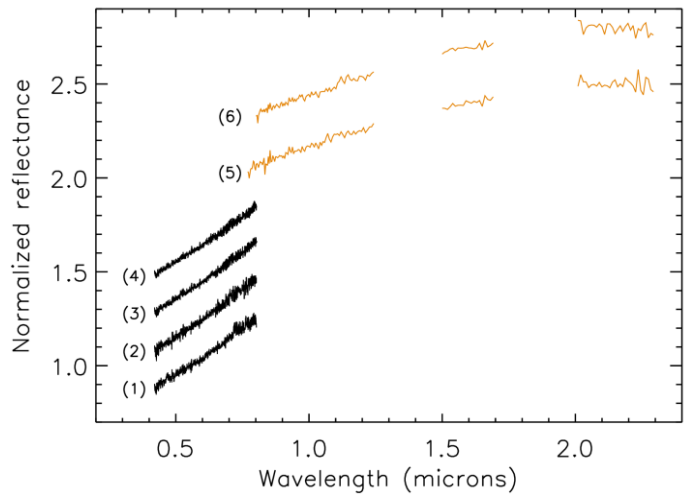

Fig. 2. Reflectance spectra of Agamemnon, shifted for clarity. Regions affected by strong atmospheric absorption (at ≈1.4 µm and ≈1.9 µm) have been cut out.

## 3. Data analysis and results

### 3.1. Search for cometary activity

High-SNR, R-filter images of Hektor and Agamemnon have been used to search for the presence of a faint coma around the targets (the R filter has been selected as the most suitable to detect any faint dust activity around the nucleus, eventually driven by residual low-level ice sublimation). In all the images, both Hektor and Agamemnon appear inactive, with a stellar appearance compared to in-field stars. To obtain a more stringent comparison, we performed an analysis of the surface brightness profile of the two targets, computed by azimuthally averaging the signal within concentric annuli around the optocentre. This is done in order to obtain the surface brightness profile (SBP) $\Sigma_\lambda(\rho)$ (given in ADU and computed for a unit solid angle [arcsec$^2$]) at various aperture radii $\rho$ [arcsec]. The SBP of Hektor and Agamemnon is compared to the stellar SBP in Figs. 3 and 4 for one of the obtained R-images during the observing night (the behaviour in the other images is quite similar). For each field, the stellar SBPs reported in the figure are obtained from several in-field stars: from the analysis of corresponding calibrated images, it is possible to conclude that no clear residual dust activity is detected down to 23.5 and 23.4 R-mag/arcsec$^2$, for Hektor and Agamemnon, respectively.

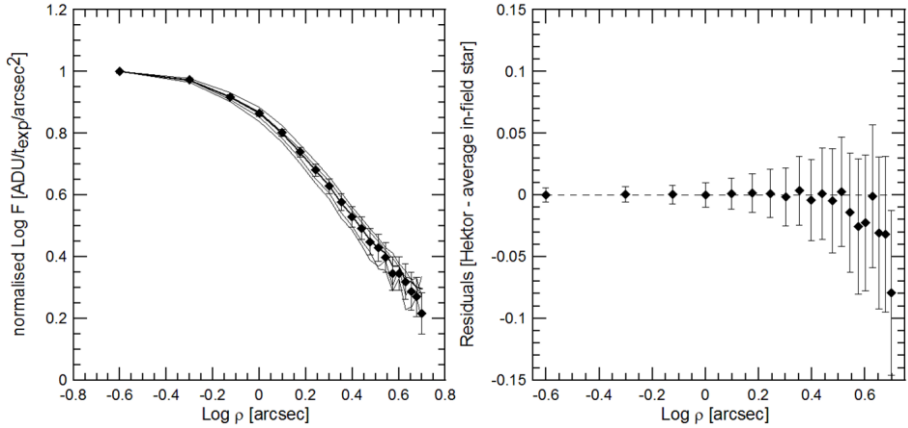

Fig. 3. (Left) comparison of SBP for Hektor and 4 in-field stars. (Right) Corresponding residuals with the average in-field stellar SBPs.

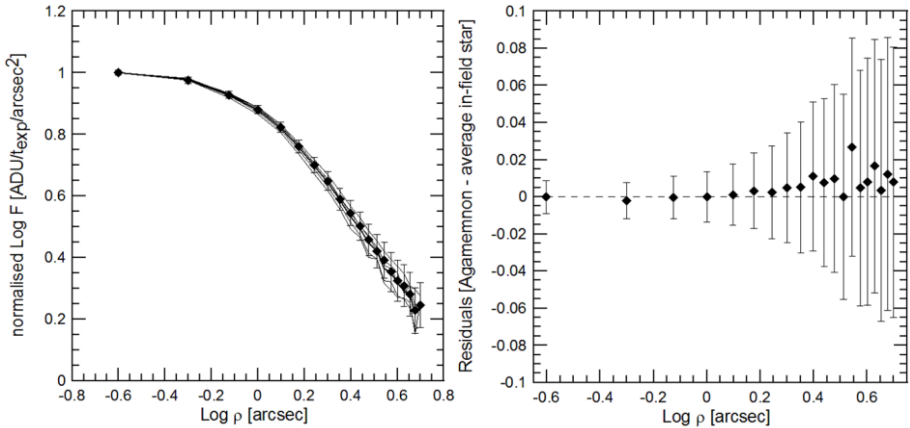

Fig. 4. (Left) comparison of SBP for Agamemnon and 5 in-field stars. (Right) Corresponding residuals with the average in-field stellar SBPs.

The possible contamination due to a weak, unresolved dust coma around the targets can be quantified by the application of a method that constrains the coma magnitude. Jewitt (1991) showed that for a standard steady-state coma, the integrated coma magnitude $m_{coma}$ within a diaphragm of radius $\rho$ is related to the SBP at $\rho$ as

$$m_{coma} = SBP - 2.5 \log(2\pi\rho^2)$$

In the case of stellar appearance, as it is for our targets, the approximation of a steady-state coma can be used to give an estimate of the amount of a possible coma contamination to the bare nucleus. This can be carried out by considering that

$$\frac{F_{coma}}{F_{TOT}} = 10^{-0.4(m_{coma} - m_{TOT})}$$

At the reference aperture of $\rho = 3.5''$ (approximately 3 × PSF), we have $m_{TOT} = 13.989$ and $m_{TOT} = 14.732$, and SBP = 21.481 and SBP = 21.939 mag/arcsec$^2$, for Hektor and Agamemnon, respectively. This results in $\frac{F_{coma}}{F_{TOT}} = 0.08$ and $\frac{F_{coma}}{F_{TOT}} = 0.10$ for Hektor and Agamemenon, respectively, i.e. the possible coma contamination is less that ~ 10% for both the Trojans.

To derive an estimate of the possible dust production rate $Q_d$ by the two Trojans, we applied a "photometric method" to relate the optical photometry to $Q_d$ via some realistic assumptions about the comet-like dust environment, in particular on the power-law dust size distribution and the dust outflow velocity. This method has been derived by the technique used by Jewitt (2009) to compute the dust production rate of active Centaurs, and has been successfully applied by our group to several minor bodies (Mazzotta Epifani et al. 2010, 2014a, 2014b, 2017; Mazzotta Epifani & Palumbo 2011). Details on the realistic assumptions needed on albedo, dust distribution and size can be found in the above papers. Here we spend only a few words on the issue of the grain velocity, which is critical due to its dependency on the dust size considered in the model. For spherical grains emitted by a perfectly homogeneous nucleus, a dust velocity close to 10% of the gas velocity is predicted by the Probstein's theory (Probstein 1969; Fulle 1990). Biver et al. (2002) showed that gas velocity follows the law $v(r) = v_0 [r_0/r]^{1/4}$ with $v_0 = 550$ m/s and $r_0 = 5$ AU (obtained from measurements of expansion speeds in comet C/1995 O1 Hale-Bopp), which would lead to a grain velocity ~50 m/s for JTs. On the other hand, numerical simulations performed to model the dust environment of Centaur P/2004 A1 (LONEOS) at ~5.5 AU (Mazzotta Epifani et al. 2006) depicted a scenario of very slow ($v$ ~ 1 m/s) rather large (~1 cm) dust grains. Actually, coupling efficiency of dust grains with gas is strictly depending on a number of factors (mainly grain size, but also dust-to-gas ratio by mass, active fraction of nucleus surface, gas production rate, etc.). We therefore adopted a dust outflow velocity range of $v = 1 - 50$ m/s for an average compact (bulk density = 1000 kg/m$^3$) dust scatterer with radius $<a> = 3$ μm (considering a power-law size distribution with power $q = -3.5$ for grains between $a_- = 0.01$ μm and $a_+ = 1$ mm). Adopting a dust albedo $A = 0.04$, the obtained photometry among the coma radius of $\phi' = 3.5''$ and $\phi'' = 4.5''$ results in a range for $Q_d$ of $1 - 30$ kg/s and $1 - 24$ kg/s for Hektor and Agamemnon, respectively. These values should in any case be regarded with caution: as strictly dependant on the adopted combination of dust size and velocity, they only give an idea of the order of magnitude of the possible undetected dust loss rate for the targets. E.g., for smaller grains ($<a> = 1$ μm) a velocity $v$ = 50 m/s would result in $Q_d$ values of 10 kg/s and 7 kg/s for Hektor and Agamemnon, respectively.

**3.2. Spectral models**

For both targets, the obtained spectra (covering about half of the surface of each object) look very similar, with no clear indications of possible absorption bands. The computed spectral slopes in the 0.55-0.80 µm, 0.90-1.65 µm and 1.65-2.20 µm ranges are reported in Table 1, together with the rotational phase at the moment of the mid-observation (an arbitrary zero-point has been assumed at the acquisition of the first spectrum of each target).

In order to put some constraints on their possible surface composition, we performed spectral modelling of the complete visible and NIR spectral range for Hektor and Agamemnon. As we did not find any hint of rotational variability, we combined the four visible spectra and the two NIR spectra obtained for each object. The final combined spectra were then normalized to the values of the visible albedo at 0.55 µm: 0.03±0.01 for Hektor (Cruikshank 1977) and 0.042±0.005 for Agamemnon (Shevchenko, Slyusarev & Belskaya 2014). Here we stress that while the NEOWISE survey more recently measured an unexpectedly high albedo for Hektor (0.107±0.011; Grav et al. 2012), a number of previous results are in agreement with the reference value that we decided to assume (Hartmann & Cruikshank 1978, 1980; Fernández, Sheppard & Jewitt 2003; Emery et al. 2006).

Spectral models were produced using the radiative transfer formalism developed by Shkuratov et al. (1999). This model uses the optical constants of individual components and gives the albedo of a particulate surface. We used the optical constants of materials that are expected to be present on the surface of JTs, such as dark carbonaceous compounds, minerals, organic materials and water ice. Table 2 contains references for all of the optical constants that we used during the modelling process. We note that optical laboratory spectra depend on conditions under which they were obtained, and on the temperature in particular. Therefore, in order to retrieve more reliable results, we used optical constants that were obtained at temperatures as close as possible to that of our target JTs (T~100K). We varied the grain size (by steps of 5 µm between 5 µm and 1 cm) and the relative abundances of each considered possible constituent (by steps of 5% for complete spectral models, and by steps of 1% for the 2-µm water ice band region, see below), and looked for the best-fitting spectral models by minimization of the root-mean-square deviation. Some successful models for Hektor and Agamemnon are presented in Figs. 5 and 6. Our final synthetic spectra are the combination of amorphous carbon, kerogen, pyroxene and water ice. We stress that even if spectral modelling is the best tool to interpret the reflectance spectra and thus to access the surface properties of a target, it is impossible to produce a unique solution. This is particularly true for featureless spectra such as those here analysed, for which we can only put some rough constraints on the concentration of the mixture components, as discussed in the following.

Both the investigated objects have very low albedo, typical of D-type asteroids. Consequently, some very dark material should be included in the spectral models (dark carbon allotropes are typically used for such cases). Then, the general spectral shape of D-type asteroids can be fitted with silicate minerals and the addition of some reddening material. In our models, for both Hektor and Agamemnon, the general shape of the spectrum is well described using pyroxene containing different amounts of magnesium (Mg-rich pyroxene tends to be redder). As it was suggested by

Gradie & Veverka (1980) and then widely accepted, dark red surfaces of JTs can be well described with red opaque polymer-type organic compounds such as kerogen or tholins. In our models, we noticed that the addition of tholins did not significantly improve the fit for both the targets. Addition of kerogen, on the other hand, helped us to produce a quite good fit to the observed spectra. However, while in our final models we do not consider tholins, we cannot completely rule out their presence on the investigated surfaces, also considering that our data do not give us access to their 3-µm absorption band.

Water ice absorption bands are absent in the spectra of Hektor and Agamemnon. Nevertheless, we included water ice in our modelling in order to estimate the upper limit for its presence on their surfaces. The strongest water ice absorption band available for detection within the spectral range of our data is at 2.03 µm. For this reason, we paid special attention to this part of the spectra. Figs. 7 and 8 present spectral models with different percentages of water ice inclusions. The depth of the water absorption band depends on both the grain size and concentration in the mixture. A strongest band can be achieved with the use of larger grains as well as larger concentrations, and vice versa. On the other hand, both grain size and concentration are mutually interchangeable, i.e. the same synthetic spectra can be obtained with different model parameters. Hence, to establish the upper limit of water ice on the surfaces, we took into account models with relatively small water ice grains in our final models. Accordingly to what above, we can roughly define an upper limit of about 3-5% for the water ice abundance.

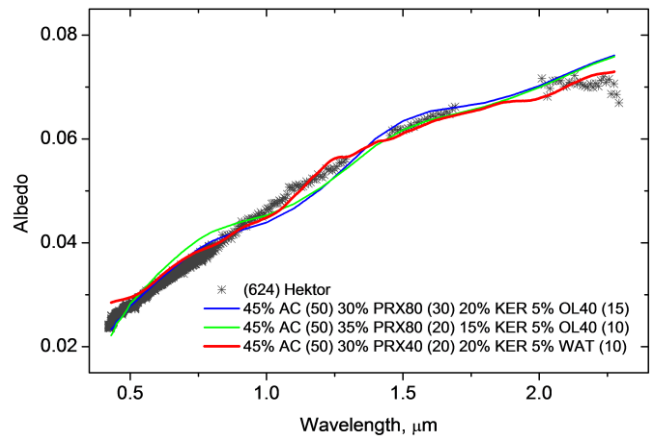

Fig. 5. Visible-NIR spectrum of Hektor and some best-fitting spectral models. The grain size is given in parentheses (µm).

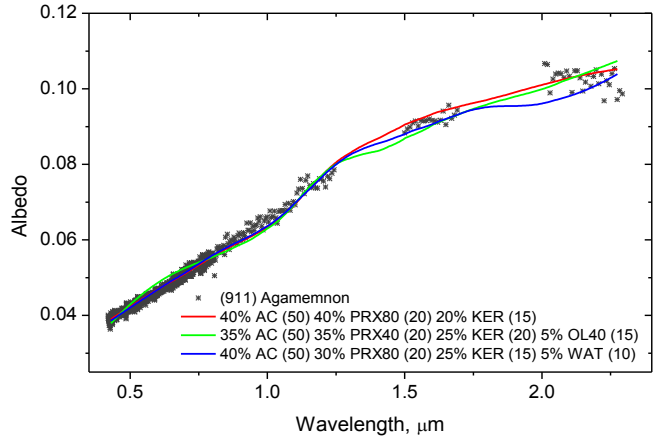

Fig. 6. Visible-NIR spectrum of Agamemnon and some best-fitting spectral models. The grain size is given in parentheses (μm).

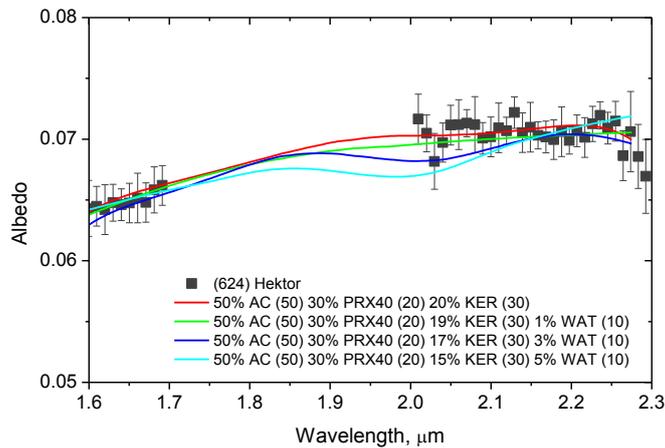

Fig. 7. Zoomed part of the NIR spectrum of Hektor, and spectral models with different concentrations of water ice. The grain size is given in parentheses (μm).

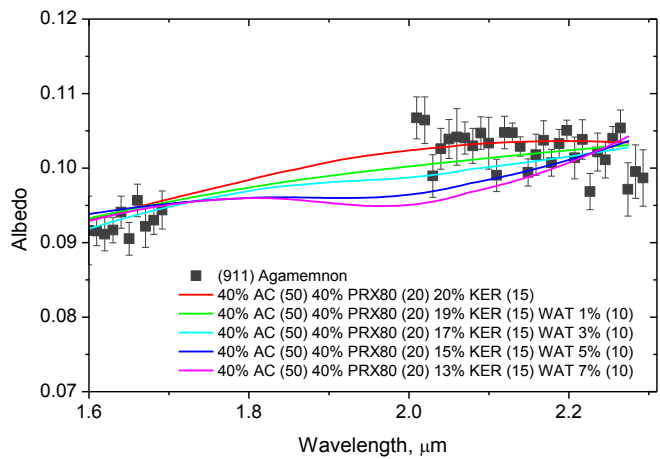

Fig. 8. Zoomed part of the NIR spectrum of Agamemnon, and spectral models with different concentrations of water ice. The grain size is given in parentheses (μm).

*Table 2. Optical constants considered.*

| Component | Acronym | Reference |
|---|---|---|
| Olivine40 ($Mg_{0.8}Fe_{1.2}SiO_4$) | OL40 | Dorschner et al. (1995) |
| Pyroxene40 ($Mg_{0.4}Fe_{0.6}SiO_3$) | PRX40 | Dorschner et al. (1995) |
| Pyroxene80 ($Mg_{0.8}Fe_{0.2}SiO_3$) | PRX80 | Dorschner et al. (1995) |
| Titan tholin | Tit. th. | Khare et al. (1994) |
| Ice tholin | Ice th. | Khare et al. (1993) |
| Amorphous carbon | AC | Zubko et al. (1996) |
| Kerogen | KER | Khare et al. (1990) |
| Water Ice | WAT | Warren (1984) <br> Grundy & Schmitt (1998) |
| Basalt | BAS | Pollack, Toon & Khare (1973) |

**Conclusions**

In this work we present new visible and near-infrared spectral observations of two of the very largest Jupiter Trojans, (624) Hektor and (911) Agamemnon. For both objects, our dataset represents the first ever acquired at different rotational phases (covering about 50% of the surface of each target). Contrarily to what spectral observations in the literature could suggest, we have found hints of extremely homogeneous surfaces (we stress that during the same observational run at the TNG facility, the dwarf planet Ceres evidenced a remarkable short-term spectral variation with the rotational phase, see Perna et al. 2015. The homogeneity of the spectral data that we present here for Hektor and Agamemnon incidentally confirms the physical nature of the variations we had found for Ceres). Moreover, we did not find any water-related absorption feature, nor sign of coma/outgassing from the analysis of the complementary photometric data that we have acquired (with upper limits of the order of ≈10 kg/s for the dust production rate, if any). Overall, our spectral modelling (using the formalism by Shkuratov et al. 1999) of both Hektor and Agamemnon is in agreement with previous results in the literature, with best-fit models including amorphous carbon, magnesium-rich pyroxene and kerogen. We can put an upper limit of a few percent to the amount of water ice on their surfaces.

      Considering the striking similarities that have been evidenced (e.g., Kelley et al. 2017) between Hektor and Agamemnon (among other Jupiter Trojans) and the cometary nuclei, such Jupiter Trojans could be considered as dead or dormant (giant) comets. Given the large size of these two bodies, they are as old as our solar system. Their homogeneous surfaces could be a result of space weathering, past cometary activity and the formation of a surface mantle. We remind the reader that, according to the latest dynamical models (Nesvorný et al. 2013), Jupiters Trojans could have spent about $10^3$-$10^4$ years with perihelion distances in the range 1.5-3 AU at the time of Jupiter migration. Outgassing could have been intense at that stage. Noteworthy, the exploration of comet 67P/Churyumov-Gerasimenko by the ESA Rosetta mission revealed variations in the spectral slope between active and inactive regions (e.g., Fornasier et al. 2015, 2016; Perna et al. 2017). Hence, the lack of surface heterogeneity on Hektor and Agamemnon can put some constraints on the epoch when any cometary activity ceased, and/or on the collisional event that formed the Hektor dynamical family.

      Finally, we want to note that the Lucy space mission (chosen in 2017 by NASA in the framework of the Discovery Program for a launch in 2021) will fly-by five Jupiter Trojans between 2027 and 2033; while Hektor and Agamemnon are not among the targets that will be explored by Lucy, this mission has the potential to significantly boost our knowledge of the Jupiter Trojan population, which is in turn key to the understanding of the solar system formation and early evolution.

**Acknowledgements**

This work is based on observations made with the Italian Telescopio Nazionale Galileo (TNG) operated on the island of La Palma by the Fundación Galileo Galilei of the INAF (Istituto Nazionale di Astrofisica) at the Spanish Observatorio del Roque de los Muchachos of the Instituto de Astrofisica de Canarias. DP has received funding from the European Union's Horizon 2020 research and innovation programme under the Marie Sklodowska-Curie grant agreement n. 664931.